\documentclass[prl,twocolumn, showpacs,nofootinbib]{revtex4}

\usepackage{amstext,amsmath,amssymb}
\usepackage[dvips]{graphicx}
\usepackage{psfrag}
\usepackage{latexsym}

\newcommand{\Ref}[1]{(\ref{#1})}

\newcommand{\N}{\mathbb{N}}

\newcommand{\R}{\mathbb{R}}

\newcommand{\CC}{{ C}}
\DeclareMathOperator{\tr}{tr}

\def\be{\begin{equation}}
\def\ee{\end{equation}}
\def\bes{\begin{eqnarray}}
\def\ees{\end{eqnarray}}
\def\nn{\nonumber}
\def\arr{\rightarrow}

\def\om{\omega}

\def\w{\wedge}

\def\f{\frac}
\def\tl{\widetilde}
\def\wtl{\widetilde}

\newcommand{\extd}{{\rm d}}
\newcommand{\SU}{\mathrm{SU}}

\newcommand{\SO}{\mathrm{SO}}

\def\pp{\partial}

\def\mm{{\cal M}}

\def\ss{{\cal S}}
\def\uu{{\cal U}}
\def\tK{\tl{K}}

\def\ka{\kappa}
\def \eps{\epsilon}

\newcommand{\lalg}[1]{\mathfrak{#1}}
\newcommand{\U}{\mathrm{U}}
\newcommand{\su}{\lalg{su}}

\newcommand{\so}{\lalg{so}}

\renewcommand{\v}{\vec}

\begin{document}

\title{3d Quantum Gravity and Effective Non-Commutative Quantum Field Theory}

\author{{\bf Laurent Freidel}\footnote{lfreidel@perimeterinstitute.ca}}
\author{{\bf Etera R. Livine}\footnote{elivine@perimeterinstitute.ca}}
\affiliation{Perimeter Institute, 31 Caroline Street North
Waterloo, Ontario, Canada N2L 2Y5}
\affiliation{Laboratoire de Physique, ENS Lyon, CNRS UMR 5672, 46 All\'ee d'Italie, 69364 Lyon Cedex 07}

\begin{abstract}

\begin{center}
{\small ABSTRACT}
\end{center}

We show that the effective dynamics of matter fields coupled to 3d quantum gravity is described after integration over the gravitational degrees of
freedom by a braided non-commutative quantum field theory symmetric under a $\kappa$-deformation of the Poincar\'e
group.
\end{abstract}

\maketitle




One of the most pressing issue in quantum gravity (QG) is the semi-classical regime. In this letter, we answer this question in the context of matter coupled to 3d gravity. We show how to recover standard quantum field theory (QFT) amplitudes in the no-gravity limit and how to compute the QG corrections.

Let us consider a matter field $\phi$ coupled to gravity,
\be
Z=\int Dg\,\int D\phi \, e^{iS[\phi,g]+iS_{GR}[g]},
\ee
where $g$ is the space-time metric, $S_{GR}[g]$ the Einstein gravity action and $S[\phi,g]$ the
action defining the dynamics of $\phi$ in the metric $g$. Our goal is to integrate out the quantum gravity
fluctuations
and derive an {\it effective action} for $\phi$ taking into account the quantum gravity correction:
$$
Z=\int D\phi e^{iS_{eff}[\phi]}.
$$
We propose  to expand the $\phi$ integration into Feynman diagrams, which depend on the
``background" metric $g$ and to compute the quantum gravity effects on these Feynman diagram evaluations:
\be
Z=\sum_\Gamma C_\Gamma \int Dg\,I_\Gamma[g]\,e^{iS_{GR}[g]}
=\sum_\Gamma C_\Gamma \tl{I}_\Gamma.
\ee
Finally, we re-sum these deformed  Feynman diagrams to identify the effective action
$S_{eff}[\phi]$ taking into account the QG corrections to the matter dynamics.
Here, we prove that this program can be explicitly realized for 3d quantum gravity. The resulting effective matter
theory is a non-commutative field theory invariant under the $\kappa$-deformed Poincar\'e group. The deformation
parameter $\kappa$ is simply related to the Newton constant for gravitation $\kappa=4\pi G$.
All technical proofs can be found in \cite{PR3}.


In a first order formalism, Riemannian 3d gravity is described in term of a frame field
$e^i_\mu dx^\mu$ and a spin connection $\om^i_\mu dx^\mu$, both valued in the Lie algebra
$\so(3)$. Indices $i$ and $\mu$ run from 0 to 2. The action is defined as:
\be
\label{action}
S[e,\om]=\f{1}{16\pi G}\int e^i\w F_i[w],
\ee
where $F\equiv d\om +\om\w\om$ is the curvature tensor of the 1-form $\om$. The equations
of motion for pure gravity impose that the connection is flat and the torsion vanishes,
\be
F[\om]=0, \qquad
T[\om,e]=d_\om e=0.
\ee
This is actually a topological field theory. Particles are introduced as topological
defects \cite{jdh}. Spinless particles are source of curvature (the spin introduces
torsion):
$$
F^i[\om]=4\pi G p^i \delta(x).
$$
Outside the particle, the space-time remains flat and the particle creates a conical
singularity with deficit angle related to the particle's mass \cite{matschull}:
\be
\theta=\ka m.
\ee
This deficit angle describes the feedback of the particle on the space-time geometry.
Since $\theta$ is obviously bounded by $2\pi$, particles have a maximal allowed mass $m_{P}=(2G)^{-1}$.
Note that the Planck mass $m_P$ in 3d does not depend on the Planck constant unlike the  Planck  length
$l_P=\hbar\, m_P^{-1}\sim \hbar G$. This feature is specific to 3d QG and does not apply to the 4d theory.


The spin foam quantization of 3d gravity is given by the
Ponzano-Regge model \cite{PR0}, which was the first ever written
QG model. It is a discretization of the continuum path integral,
$Z=\int De D\om e^{iS[e,\om]}$. Since the theory is topological,
the discretization actually provides an exact quantization.
Considering a triangulation $\Delta$ of a 3d manifold $\mm$ and a
graph $\Gamma\subset\Delta$, we insert particles with deficit
angles $\theta_e$ for all edges $e\in\Gamma$ of the graph.
The partition function is defined as the product of weights associated to the edges and
to the tetrahedra:
\be
\label{Igamma}
I_{\Delta}[\Gamma]=
\sum_{\{j_e\}}\prod_{e\notin\Gamma}d_{j_e}
\prod_{e\in\Gamma}
K_{\theta_e}(j_e)
\prod_t \left\{\begin{array}{ccc}
j_{e_1}&j_{e_2}&j_{e_3} \\
j_{e_3}&j_{e_5}&j_{e_6}
\end{array}
\right\},
\ee
where we sum over all assignments of $\SO(3)$ representation
$j_e\in\N$ to the edges of $\Delta$. $d_j=(2j+1)$ is the dimension
of the $j$-representation and we associate a $\{6j\}$ symbol to
each tetrahedron. $h_\theta=\exp(i\theta \sigma_3)$ is in the
$\U(1)$ subgroup and we define the weight:
$$
K_\theta(j)=\f{i}{2\ka^2}\f{e^{-id_j(\theta-i\eps)}}{\cos\theta},
\quad
{\rm Re}\,K_\theta= \f{\cos\theta}{2\ka^2\sin\theta}\,\chi_j(\theta)
$$
where $\eps>0$ is a regulator and $\chi_j(\theta)$ the trace of $h_\theta$ in the
$j$-representation. $K_\theta$ defines the insertion of a Feynman propagator while ${\rm
Re}\,K_\theta$ gives a Hadamard propagator and leads back to the same partition function
as in
\cite{PR1}.

The partition function has a dual formulation in terms of $\SO(3)$ group elements
attached to the faces $f\in\Delta$:
\bes
&&I_\Delta[\Gamma,\theta_e]=
\int \prod_f \extd g_f\,
\prod_{e\in\Gamma} \tK_{\theta_e}(g_e)\,
\prod_{e\notin\Gamma} \delta(g_e),
\label{Igamma2} \\
&& \tK_{\theta_e}(h_\phi)=
\f{i}{\ka^2}\f{1}{(\sin^2\phi -\sin^2\theta_e + i\eps)}
=\sum_j K_\theta(j)\chi_j(\phi), \nonumber
\ees
where $g_e$ is the oriented product $\prod_{\pp f\ni e} g_f$ and the function
$\tK_\theta(g)$ is invariant under conjugation. Using the real part of $K_\theta$ leads
to replacing $\tK_\theta$ by the distribution $\delta_\theta(g)$ which fixes the rotation
angle of $g$ to $\theta$,
$$
\int_{\SO(3)} \extd g f(g) \delta_\theta(g)=
\int_{\SO(3)/\U(1)} \extd x f(xh_\theta x^{-1}).
$$

$I_\Delta[\Gamma]$ is independent of the triangulation $\Delta$ and only depends on the topology of $(\mm,\Gamma)$.
It is finite after suitable gauge fixing of the diffeomorphism symmetry \cite{diffeo}, which removes redundancies in
 the product of $\delta$-functions. Then for a trivial topology $\mm=[0,1]\times\Sigma_2$, $I_\Delta$
is the projector onto the physical states, that is the space the flat connections on $\Sigma_2$  \cite{ooguri}.
Moreover, this quantization scheme has been shown to be equivalent to the Chern-Simons quantization \cite{PR2}.
Finally, the large $j$ asymptotics of the $\{6j\}$ symbols are related to the discrete Regge action for 3d gravity \cite{6jasymp}.

We have defined a purely algebraic quantum gravity amplitude $I_\Delta[\Gamma]$. The Newton constant $G$ only
appears as a unit to translate the algebraic quantities $j,\theta$ into the physical length $l=jl_P=j\hbar G$ and the
physical mass $m=\theta/\ka=\theta/4\pi G$.


The essential point is that the QG amplitudes $I_\Delta[\Gamma]$ are the Feynman diagram evaluations of a
 non-commutative field theory. Let us first consider a trivial topology
$\mm\sim \ss^3$ with $\Gamma$ planar. In this case, we can get rid of the triangulation dependence and rewrite
 $I_{\Gamma}\equiv I_{\Delta}[\Gamma]$ as\cite{PR3}:
\be\label{Igamma3}
{I}_{\Gamma}= \int\prod_{v\in\Gamma} \frac{\extd^3 X_v}{8\pi\ka^3}\,
\int\prod_{e\in \Gamma} \extd g_e\,
\tK_{\theta_e}(g_e)\,
\prod_{v \in \Gamma}e^{\frac{1}{2\ka}\,\tr(X_vG_v)}.
\ee
The integral is over one copy of $\so(3)\sim\R^3$ for each vertex $X_{v}\equiv X_{v}^{i}\sigma_{i}$ and one copy
of $\SO(3)$ for each edge. We define   at each vertex $v$, the ordered product of the edge group elements meeting at $v$
\be
G_v=\overrightarrow{\prod_{e\supset v}}g_e^{\epsilon_v(e)},
\ee
$\epsilon_v(e)=\pm 1$ depending on whether the edge is incoming or outgoing at $v$. The
kernel $\tK_\theta$ defines the Feynman propagator and is given by
\be \label{prop}
\tK_\theta(g)=\int_{\R^+} {dT} e^{iT\left(P^2(g) -\left(\frac{\sin{\ka
m}}{\ka}\right)^2\right)},
\ee
with  $2i\vec{P}(g)\equiv \tr(g\vec{\sigma})$ the projection  of $g$ on  Pauli matrices.
Changing the integration range from $\R^+$ to $\R$, we would obtain the Hadamard function
$\delta_\theta$ instead of $\tK_\theta$.
To further simplify this expression, we introduce a noncommutative $\star$-product on
$\R^3$ such that
\be\label{star}
e^{\frac{1}{2\ka}\mathrm{\tr}(Xg_{1})}\star
e^{\frac{1}{2\ka}\mathrm{\tr}(Xg_{2})}
 = e^{\frac{1}{2\ka}\mathrm{\tr}(Xg_{1}g_{2})},
\ee
Using the parametrization of $\SO(3)$ group elements,
\be
g= (P_4  + \imath \ka P^i\sigma_i),
\quad P_4^2 +\ka^2 P^iP_i=1,
\quad P_{4}\geq 0,
\nn
\ee
the $\star$-product deforms the composition of plane waves,
\bes
 e^{\imath (\vec{P}_1
\oplus \vec{P}_2) \cdot \vec{X}}&=&e^{\imath\vec{P}_1\cdot \vec{X}}
\star e^{\imath \vec{P}_2\cdot \vec{X}}, \label{starP} \\
\vec{P}_1 \oplus \vec{P}_2 &= &\sqrt{1-\ka^2|\vec
P_2|^2}\,\vec P_1 + \sqrt{1-\ka^2|\vec P_1|^2}\, \vec P_2 \nn\\
&& - \ka\vec P_1\times \vec P_2,
\label{oplus}
\ees
with $\times$ the 3d vector cross product.
To define the $\star$-product on all functions, we introduce a new {\it group Fourier transform}
$F: C(\SO(3))\to \CC_{\ka}(\R^3)$ mapping functions on the group $\SO(3)$ to functions on $\R^3$
with momenta bounded  by $1/\ka$:
\be
\phi(X)= \int \extd g \,\tilde{\phi}(g) e^{\frac{1}{2\ka}\mathrm{\tr}(Xg)}.
\ee
The inverse group Fourier transform  is explicitly written
\bes
\tilde\phi(g)&=&\int_{\R^3} \frac{d^3X}{8\pi \ka^3}\,\, \phi(X)
\star e^{\frac{1}{2\ka} \tr(Xg^{-1})}\\
&=&\int_{\R^3} \frac{\extd^3X}{8\pi \ka^3}\,\, \phi(X)\nonumber
\sqrt{1-\ka^2 P^2(g)} e^{\frac{1}{2\ka} \tr(Xg^{-1})}.
\ees
Under this Fourier transform, the $\star$-product is dual to the group convolution product.
Finally $F$ is an isometry between $L^2(\SO(3))$ and $\CC_{\ka}(\R^3)$ equipped with the norm
\be
||\phi||_{\ka}^2= \int \frac{\extd X}{8\pi \ka^3}\, \phi\star\phi(X).
\ee
Using this $\star$-product, the amplitude (\ref{Igamma3}) reads
\be\label{Igamma4}
{I}_{\Gamma}=  \int\prod_{v\in\Gamma} \frac{\extd X_v}{8\pi\ka^3}\,
\prod_{e\in \Gamma} \extd g_e\,\tK_{\theta_e}(g_e)\,
\prod_{v \in \Gamma}\left(\underset{ v\in \Gamma}{\bigstar}
e^{\frac{\epsilon_v(e) }{2\ka}\,\mathrm{\tr}(X_vg_e)} \right).
\ee
Let us now restrict to the case where we have particles of only
one type so all masses are taken equal, $m_e\equiv m$ and  consider the sum over trivalent graphs:
\be
\sum_{\Gamma {\mathrm{trivalent}}}
\f{\lambda^{|v_\Gamma|}}{S_\Gamma} {I}_\Gamma
\ee
where $\lambda$ is a coupling constant. $|v_\Gamma|$ is the number
of vertices of $\Gamma$ and $S_\Gamma$ is the symmetry factor of
the graph. Remarkably, this sum can be obtained from the
perturbative expansion of a non-commutative field theory given
explicitly by:
\bes
\nn
S=\int \f{d^3x}{8\pi\ka^3} &&\left[\f{1}{2}(\partial_i\phi \star \partial_i\phi)(x) -\f{1}{2}\f{\sin^2
m\ka}{\ka^2} (\phi \star \phi)(x)\right.\\
  &&+ \left.\f{\lambda}{3!} (\phi \star \phi\star \phi)(x)
\right]
\label{Seff}
\ees
where the field $\phi$ is in ${\CC}_\ka(\R^3)$.
Its momentum has support in the ball of radius $\ka^{-1}$.
We can write this action in momentum space
\bes
S(\phi)&=&
\f{1}{2}\int dg\, \left(P^2(g)- \f{\sin^2\ka m}{\ka^2}\right)\wtl{\phi}(g)\wtl{\phi}(g^{-1}) \\ \nonumber
&+&\f{\lambda}{3!}\int dg_1dg_2 dg_3 \,\delta(g_1g_2g_3)\,\wtl{\phi}(g_1)\wtl{\phi}(g_2)\wtl{\phi}(g_3).
\label{Seffmom}
\ees
This is our effective field theory describing the dynamics of the
matter field after integrating out the gravitational sector.
This non-commutative field theory action is symmetric under a $\kappa$-deformed action of the Poincar\'e group.
Calling $\Lambda$ the generators of Lorentz transformations and $T_{\vec{a}}$ the generators of translations,
the action of these generators on one-particle states is undeformed:
\bes
\label{1paction}
\Lambda\cdot \tl{\phi}(g) &=& \tl{\phi}(\Lambda g \Lambda^{-1})= \tl{\phi}(\Lambda \cdot P(g)), \\
T_{\vec{a}}\cdot\tl{\phi}(g) &=& e^{i\vec{P}(g)\cdot\vec{a}}\tl{\phi}(g).
\ees
The non-trivial deformation of the Poincar\'e group appears at the level of
multi-particle states and only the action of the translations is deformed :
\bes
\label{2paction}
\Lambda \cdot\tl{\phi}(P_1) \tl{\phi}(P_2)&=& \tl{\phi}(\Lambda \cdot P_1)\tl{\phi}(\Lambda \cdot P_2), \\
T_{\vec{a}}\cdot\tl{\phi}(P_1) \tl{\phi}(P_2) &=& e^{i\vec{P}_1\oplus \vec{P}_2\cdot\vec{a}}\tl{\phi}(P_1)\tl{\phi}(P_2).
\ees

\begin{figure}[t]
\hspace{-3cm}
\psfrag{1}{$g_1$}
\psfrag{2}{$g_2$}
\psfrag{3}{$g_3$}
\psfrag{b}{$g'_1$}
\psfrag{a}{$g'_2$}
\psfrag{V}{$\equiv \delta(g_1g_2g_3)$}
\psfrag{P}{$\equiv K_m(g_1)$}
\psfrag{B}{$\equiv \delta(g_1g_2g'_1{}^{-1}g'_2{}^{-1})\,\delta(g_2g'_2{}^{-1})$}
\includegraphics[width=4.3cm]{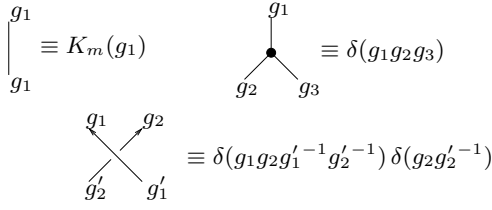}
\caption{Feynman rules for particles propagation in the Ponzano-Regge model.}
\label{Feynmrules1}
\end{figure}

It is straightforward to derive the Feynman rules from the action \Ref{Seffmom} (see fig.\ref{Feynmrules1}).
The effective Feynman propagator is the group Fourier transform of $\tK_\theta(g)$,
\be
K_{m}(X) =i\int \extd g \frac{e^{\frac{1}{2\ka}\,\mathrm{\tr}(X g)} }{P^2(g) -
\left(\frac{\sin{\ka m}}{\ka}\right)^2}.
\ee
The effect of quantum gravity is two-fold. First the mass gets renormalized  $m\to{\sin{\ka m}}/{\ka} $.
Then the momentum space is no longer the flat space but the homogeneously curved space $S^{3}\sim\SO(3)$.
This reflects that the momentum is bounded $|P| <1/\ka$.

\noindent
At the interaction vertex the momentum addition becomes non-linear with a conservation
rule $P_1\oplus P_2 \oplus P_3=0$ which implies a non-conservation of momentum
$P_1+P_2+P_3\ne 0$. Intuitively, part of the energy involved in a collision process is
absorbed by the gravitational field: gravitational effects can not be ignored at high
energy. This effect, which is stronger at high momenta and for non-collinear momenta,
prevents the total momenta from being larger than the Planck energy.

\noindent
A last subtlety of the Feynman rules is the evaluation of non-planar diagrams.
A careful analysis of $I_\Gamma$ shows that we have a non-trivial {\it braiding}:
for each crossing of two edges, we associate a weight $\delta(g_1g_2g'_1{}^{-1}g'_2{}^{-1})\,\delta(g_2g'_2{}^{-1})$
(see fig.\ref{Feynmrules1}).
This reflects a non-trivial statistics where the Fourier modes of the fields
obey the exchange relation:
\be\label{stat}
\tl{\phi}(g_1)\tl{\phi}(g_2)=
\tl{\phi}(g_2)\tl{\phi}(g_2^{-1}g_1g_2)
\ee
which is naturally determined by our choice of star pro\-duct. Indeed, let us look at the product of two identical fields:
\be
\phi\star \phi\,(X) = \int dg_1 dg_2\,
 e^{\f{1}{2\ka} \tr(Xg_1g_2)} \wtl{\phi}(g_1)\wtl{\phi}(g_2),
\ee
Under change of variables $(g_1,g_2)\rightarrow (g_2,g_2^{-1}g_1g_2)$, the star product reads
\be
\phi\star \phi\,(X) = \int dg_1 dg_2\,
 e^{\f{1}{2\ka} \tr(Xg_1g_2)} \wtl{\phi}(g_2)\wtl{\phi}(g_2^{-1}g_1g_2).
\ee
The identification of the Fourier modes of $\phi\star \phi\,(X)$ leads to the exchange relation \Ref{stat}.
This braiding was first proposed in \cite{schroers} for two particles coupled to 3d QG and then computed in the Ponzano-Regge model in \cite{PR1}.
It is encoded into a braiding matrix
\be
R\cdot\tl{\phi}(g_1)\tl{\phi}(g_2)=
\tl{\phi}(g_2)\tl{\phi}(g_2^{-1}g_1g_2).
\ee
This is the $R$ matrix of the $\kappa$-deformation of the Poincar\'e group \cite{schroers}.
Such field theories with non-trivial braided statistics are simply called braided non-commutative field
theories and were first introduced in \cite{Robert}.

Finally, the $\star$-product induces a non-commutativity of space-time and a deformation of phase space:
\begin{eqnarray}
{[}X_i,X_j{]}&=&i\ka\epsilon_{ijk} X_k, \nonumber\\
{[}X_i,P_j{]}&=&i\sqrt{1-\ka^2 P^2}\,\delta_{ij}-i\ka\epsilon_{ijk}P_k.
\end{eqnarray}
This non-commutativity reflects the fact that momentum space is curved. Indeed the
coordinates $X$ are realized as right invariant derivations on momentum space and
derivations of a curved manifold do not commute. Moreover, this non-commutativity being
related to having bounded momenta implies the existence of a minimal length scale
accessible in the theory. Indeed defining the non-commutative $\delta$-function
$\delta_{0}\star \phi (X)= \phi(0)
\delta_{0}(X)$, we compute
\be
\delta_{0}(X)= 2\ka\frac{J_{1}\left(\frac{|X|}{\ka}\right)}{|X|},
\ee
with $J_{1}$ the $1$st Bessel function. It is clear that $\delta_{0}(X)$ is concentrated  around $X=0$
but has a non-zero width.


Using this formalism, one can compute the QG effects order by order in $\ka$.
The 0th order is defined by the no-gravity limit $\ka\arr 0$.
Starting either from the spinfoam amplitude $I_\Gamma$ given by \Ref{Igamma} or from the Feynman evaluations
\Ref{Igamma4}, one can show that the limit $\ka\arr 0$ is exactly given by the Feynman evaluations of the usual
commutative QFT:
\bes
\label{abelian}
&&I^0_\Delta[\Gamma,m_e]\equiv
\lim_{\ka\arr 0} \ka^{3|e_\Gamma|}I_\Delta[\Gamma,\theta_e]\\
&&=\int_{\R^3}\prod_{f\in\Delta}d^3\v{p}_f
\prod_{e\in\Gamma}\f{i}{2\pi(p_e^2-m_e^2)}
\prod_{e\in\Delta\setminus\Gamma}\delta(\v{p}_e),\nn
\ees
where $|e_\Gamma|$ is the number of edges of the graph $\Gamma$, $\v{p}_f\in\R^3$ are
variables attached to the faces of $\Delta$ and $\v{p}_e=\sum_{f\supset e} \v{p}_f$.
Moreover, since physical lengths and masses are defined in $\ka$-units, $l=\ka j$ and
$m=\theta/\ka$, taking $\ka\arr0$ corresponds to $j\arr \infty$ and $\theta\arr 0$. For
$\theta\sim 0$, the group multiplication on $\SO(3)$ becomes abelian at first order in
$\ka$. More precisely, we prove in \cite{PR3} that the no-gravity limit of the
Ponzano-Regge model is actually the topological state sum based on the abelian group
$\R^3$.
This shows that the usual Feynman evaluations of QFT in 3d can be generically written as
amplitudes of a topological theory.


Up to now, we have worked in the Riemannian context. All the
previous constructions and results can be straightforwardly
extended to the Lorentzian theory.
The Lorentzian version of the Ponzano-Regge model is expressed in
terms of the $\{6j\}$ symbols of the non-compact group $\SO(2,1)$
\cite{lorentz1}. Holonomies around particles are $\SO(2,1)$ group
elements parametrized as $g=P_4 +i\ka P_i \tau^i$ with
$P_4^2+\ka^2P_iP^i=1$ and $P_4\ge0$,
with the metric $(+--)$ and the $\su(1,1)$ Pauli matrices,
$\tau_0=\sigma_0, \tau_{1,2}=i\sigma_{1,2}$. Massive particles
correspond to the $P_iP^i>0$ sector. They are described by
elliptic group elements, $P_4=\cos\theta$, $\ka |P| =\sin\theta$.
The deficit angle is given by the mass, $\theta=\ka m$. All the
mathematical relations of the Riemannian theory are translated to
the Lorentzian framework by changing the signature of the metric.
The propagator remains given by the formula \Ref{prop}.
The momentum space is now ${\rm AdS}^3\sim\SO(2,1)$. The addition
of momenta is deformed accordingly to the formula \Ref{oplus}. We
similarly introduce a group Fourier transform $F:C(\SO(2,1))\arr
C_\ka(\R^3)$ and a $\star$-product dual to the convolution product
on $\SO(2,1)$.
Finally we derive the effective non-commutative field theory with the same expression \Ref{Seff} as in the Riemannian
case.


To sum up, we have shown that the 3d quantum gravity amplitudes, defined through the
Ponzano-Regge spinfoam model, are actually the Feynman diagram evaluations of a (braided)
non-commutative QFT. This effective field theory describes the dynamics of the matter
field after integration of the gravitational degrees of freedom. The theory is invariant
under a $\kappa$-deformation of the Poincar\'e algebra, which acts non-trivially on
many-particle states. This is an explicit realization of a QFT in the framework of
deformed special relativity (see e.g. \cite{DSR}), which implements from first principles
the original idea of Snyder of using a curved momentum space to regularize the Feynman
diagrams.

The formalism can naturally take into account a non-zero cosmological constant $\Lambda$.
The model is based on $\uu_q(\SU(2))$ and its Feynman rules are given in
\cite{PR3}.

A natural  question concerns the unitarity of our non-commutative
quantum field theory since the non commutativity affects time
\cite{NonU}. We a priori do not expect a unitary theory: since we
have integrated out the gravity degrees of freedom, we expect ghosts to appear at the
Planck energy $m_P=1/\ka\sim 1/G$.





Finally the present results suggest an extension to 4d. The standard 4d QFT Feynman
graphs would be expressed as expectation values of a 4d topological spinfoam model (see
e.g. \cite{artem,aristide}). That model would provide the semi-classical limit of QG and
be identified as the zeroth order of an expansion in term of the inverse Planck mass
$\ka$ of the full QG spinfoam amplitudes. QG effects would then appear as deformations of
the Feynman graph evaluations and QG corrections to the scattering amplitudes could be
computed order by order in $\ka$.



\end{document}